\documentclass[10pt,twocolumn,letterpaper]{article}

\usepackage{wacv}
\usepackage{times}
\usepackage{epsfig}
\usepackage{graphicx}
\usepackage{amsmath}
\usepackage{amssymb}
\usepackage{amsmath,amssymb}
\usepackage{graphicx} \graphicspath{ {Images/} }
\usepackage{amsmath}
\usepackage{booktabs}
\usepackage{microtype}
\usepackage{enumitem}
\usepackage[subtle,margins=normal,leading=normal]{savetrees}

\wacvfinalcopy 

\ifwacvfinal
\def\assignedStartPage{1} 
\fi

\ifwacvfinal
\usepackage[breaklinks=true,bookmarks=false,colorlinks]{hyperref}
\else
\usepackage[pagebackref=true,breaklinks=true,colorlinks,bookmarks=false]{hyperref}
\fi

\ifwacvfinal
\else
\fi

\begin{document}

\title{UNETR: Transformers for 3D Medical Image Segmentation}

\author{Ali Hatamizadeh\\
NVIDIA\\
\and
Yucheng Tang\\
Vanderbilt University\\
\and
Vishwesh Nath\\
NVIDIA\\
\and
Dong Yang\\
NVIDIA\\
\and
Andriy Myronenko\\
NVIDIA\\
\and
Bennett Landman\\
Vanderbilt University\\
\and
Holger R. Roth\\
NVIDIA\\
\and
Daguang Xu\\
NVIDIA\\
}

\maketitle

\begin{abstract}
   Fully Convolutional Neural Networks (FCNNs) with contracting and expanding paths have shown prominence for the majority of medical image segmentation applications since the past decade. In FCNNs, the encoder plays an integral role by learning both global and local features and contextual representations which can be utilized for semantic output prediction by the decoder. Despite their success, the locality of convolutional layers in FCNNs, limits the capability of learning long-range spatial dependencies. Inspired by the recent success of transformers for Natural Language Processing (NLP) in long-range sequence learning, we reformulate the task of volumetric (3D) medical image segmentation as a sequence-to-sequence prediction problem. We introduce a novel architecture, dubbed as UNEt TRansformers (UNETR), that utilizes a transformer as the encoder to learn sequence representations of the input volume and effectively capture the global multi-scale information, while also following the successful ``U-shaped'' network design for the encoder and decoder. The transformer encoder is directly connected to a decoder via skip connections at different resolutions to compute the final semantic segmentation output. We have validated the performance of our method on the Multi Atlas Labeling Beyond The Cranial Vault (BTCV) dataset for multi-organ segmentation and the Medical Segmentation Decathlon (MSD) dataset for brain tumor and spleen segmentation tasks. Our benchmarks demonstrate new state-of-the-art performance on the BTCV leaderboard. 
   \\
   Code: \href{https://monai.io/research/unetr}{https://monai.io/research/unetr}
\end{abstract}

\section{Introduction}

Image segmentation plays an integral role in quantitative medical image analysis as it is often the first step for analysis of anatomical structures ~\cite{monteiro2020multiclass}. Since the advent of deep learning, FCNNs and in particular ``U-shaped`` encoder-decoder architectures~\cite{isensee2019attempt,jin2020ra,isensee2021nnu} have achieved state-of-the-art results in various medical semantic segmentation tasks~\cite{BratsAll2018,simpson2019large,heller2019kits19}. In a typical U-Net~\cite{Ronneberger15} architecture, the encoder is responsible for learning global contextual representations by gradually downsampling the extracted features, while the decoder upsamples the extracted representations to the input resolution for pixel/voxel-wise semantic prediction. In addition, skip connections merge the output of the encoder with decoder at different resolutions, hence allowing for recovering spatial information that is lost during downsampling. 

\begin{figure}[t]
\includegraphics[width=\linewidth]{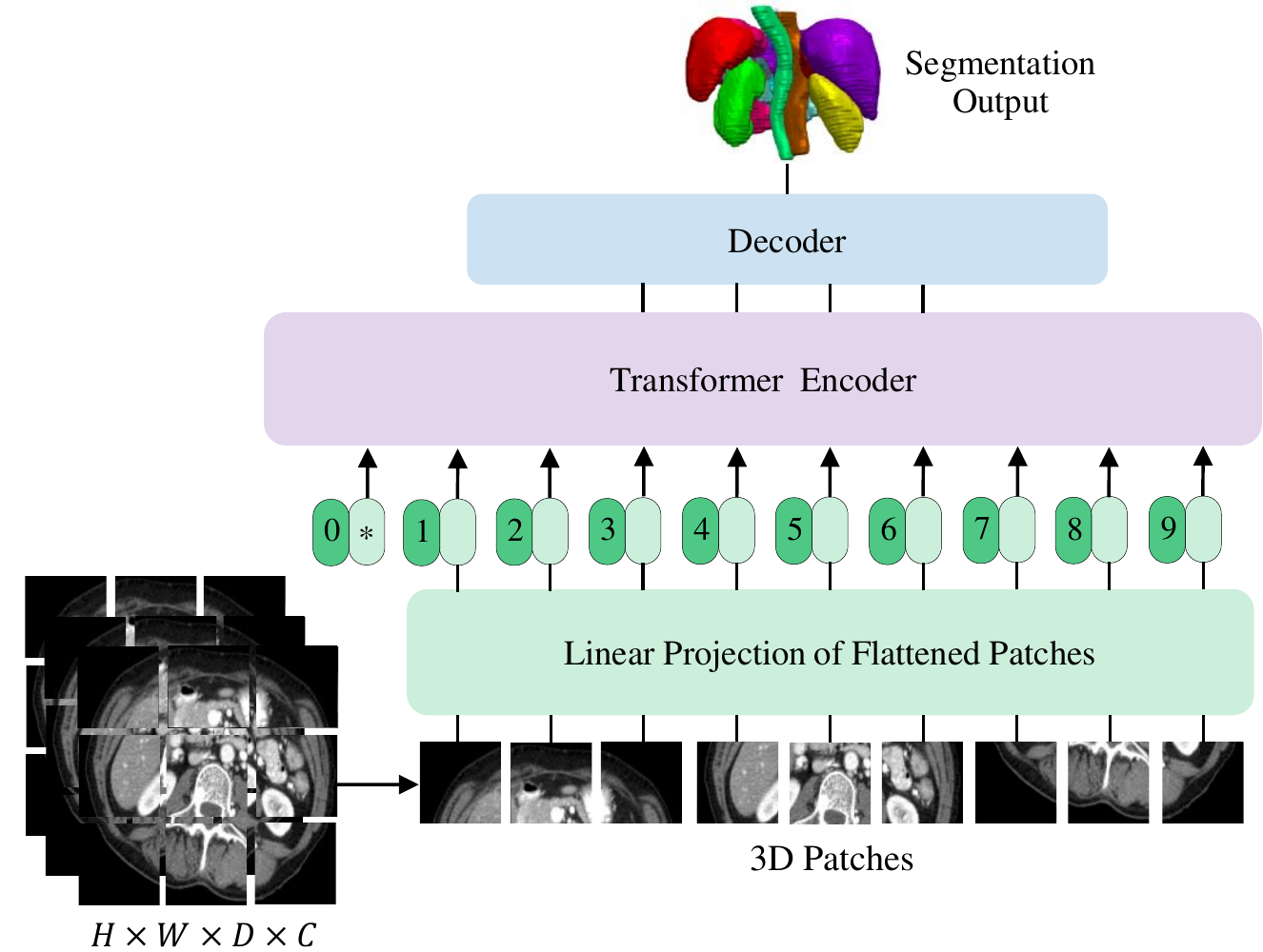}
\caption{Overview of UNETR. Our proposed model consists of a transformer encoder that directly utilizes 3D patches and is connected to a CNN-based decoder via skip connection.}
\label{fig:pipeline_highlevel}
\end{figure}

Although such FCNN-based approaches have powerful representation learning capabilities, their performance in learning long-range dependencies is limited to their localized receptive fields~\cite{hu2019local,ramachandran2019stand}. As a result, such a deficiency in capturing multi-scale information leads to sub-optimal segmentation of structures with variable shapes and scales (e.g. brain lesions with different sizes). Several efforts have used atrous convolutional layers~\cite{chen2017deeplab,li2017compactness,gu2019net} to enlarge the receptive fields. However, locality of the receptive fields in convolutional layers still limits their learning capabilities to relatively small regions. Combining self-attention modules with convolutional layers~\cite{wang2018non,zhang2019self,fu2019dual} has been proposed to improve the non-local modeling capability. 

In Natural Language Processing (NLP), transformer-based models~\cite{vaswani2017attention,devlin2018bert} achieve state-of-the-art benchmarks in various tasks. The self-attention mechanism of transformers allows to dynamically highlight the important features of word sequences. Additionally, in computer vision, using transformers as a backbone encoder is beneficial due to their great capability of modeling long-range dependencies and capturing global context~\cite{dosovitskiy2020image,bello2020lambdanetworks}. Specifically, unlike the local formulation of convolutions, transformers encode images as a sequence of 1D patch embeddings and utilize self-attention modules to learn a weighted sum of values that are calculated from hidden layers. As a result, this flexible formulation allows to effectively learn the long-range information. Furthermore, Vision Transformer (ViT)~\cite{dosovitskiy2020image} and its variants have shown excellent capabilities in learning pre-text tasks that can be transferred to down-stream applications~\cite{touvron2021training,caron2021emerging,bao2021beit}.

In this work, we propose to leverage the power of transformers for volumetric medical image segmentation and introduce a novel architecture dubbed as UNEt TRansformers (UNETR). In particular, we reformulate the task of 3D segmentation as a 1D sequence-to-sequence prediction problem and use a transformer as the encoder to learn contextual information from the embedded input patches. The extracted representations from the transformer encoder are merged with the CNN-based decoder via skip connections at multiple resolutions to predict the segmentation outputs. Instead of using transformers in the decoder, our proposed framework uses a CNN-based decoder. This is due to the fact that transformers are unable to properly capture localized information, despite their great capability of learning global information.


We validate the effectiveness of our method on 3D CT and MRI segmentation tasks using Beyond the Cranial Vault (BTCV) ~\cite{landman2015miccai} and Medical Segmentation Decathlon (MSD) ~\cite{simpson2019large} datasets. In BTCV dataset, UNETR achieves new state-of-the-art performance on both Standard and Free Competition sections on its leaderboard. UNETR outperforms the state-of-the-art methodologies on both brain tumor and spleen segmentation tasks in MSD dataset. 

our main contributions of this work are as follows::

\begin{itemize}
\item We propose a novel transformer-based model for volumetric medical image segmentation. 
\item To this end, we propose a novel architecture in which (1) a transformer encoder directly utilizes the embedded 3D volumes to effectively capture long-range dependencies; (2) a skip-connected decoder combines the extracted representations at different resolutions and predicts the segmentation output. 
\item We validate the effectiveness of our proposed model for different volumetric segmentation tasks on two public datasets: BTCV~\cite{landman2015miccai} and MSD~\cite{simpson2019large}. UNETR achieves new state-of-the-art performance on \emph{leaderboard} of BTCV dataset and outperforms competing approaches on the MSD dataset. 
\end{itemize}
\section{Related Work}


\paragraph{CNN-based Segmentation Networks}: Since the introduction of the seminal U-Net~\cite{Ronneberger15}, CNN-based networks have achieved state-of-the-art results on various 2D and 3D various medical image segmentation tasks~\cite{dou20163d,zhu2017deeply,yu2017volumetric,gibson2018automatic,li2018h}. For volume-wise segmentation, tri-planar architectures are sometimes used to combine three-view slices for each voxel, also known for 2.5D methods ~\cite{li2018h,liu20183d, xia20203d}. In contrast, 3D approaches directly utilize the full volumetric image represented by a sequence of 2D slices or modalities. The intuition of employing varying sizes was followed by multi-scan, multi-path models ~\cite{kamnitsas2015multi,kamnitsas2017efficient,chen2016combining} to capture downsampled features of the image. In addition, to exploit 3D context and to cope with limitation of computational resource, researchers investigated hierarchical frameworks. 

Some efforts proposed to extract features at multiple scales or assembled frameworks ~\cite{isensee2021nnu}. Roth~\textit{et al.}~\cite{roth2017hierarchical} proposed a multi-scale framework to obtain varying resolution information in pancreas segmentation. These methods provide pioneer studies of 3D medical image segmentation at multiple levels, which reduces problems in spatial context and low-resolution condition. Despite their success, a limitation of these networks is their poor performance in learning global context and long-range spatial dependencies, which can severely impact the segmentation performance for challenging tasks. 



\paragraph{Vision Transformers}: Vision transformers have recently gained traction for computer vision tasks. Dosovitskiy \textit{et al.}~\cite{dosovitskiy2020image} demonstrated state-of-the-art performance on image classification datasets by large-scale pre-training and fine-tuning of a pure transformer. In object detection, end-to-end transformer-based models have shown prominence on several benchmarks~\cite{carion2020end,zhu2020deformable}. Recently, hierarchical vision transformers with varying resolutions and spatial embeddings~\cite{liu2021swin,wang2021pyramid,chu2021twins,xu2021co} have been proposed. These methodologies gradually decrease the resolution of features in the transformer layers and utilize sub-sampled attention modules. Unlike these approaches, the size of representation in UNETR encoder remains fixed in all transformer layers. However, as described in Sec.~\ref{sec:method}, deconvolutional and convolutional operations are used to change the resolution of extracted features.

Recently, multiple methods were proposed  that explore the possibility of using transformer-based models for the task of 2D image segmentation ~\cite{zheng2020rethinking,chen2021transunet,valanarasu2021medical,zhang2021transfuse}. Zheng \textit{et al.}~\cite{zheng2020rethinking} introduced the SETR model in which a pre-trained transformer encoder with different variations of CNN-based decoders were proposed for the task of semantic segmentation. Chen \textit{et al.}~\cite{chen2021transunet} proposed a methodology for multi-organ segmentation by employing a transformer as an additional layer in the bottleneck of a U-Net architecture. Zhang \textit{et al.}~\cite{zhang2021transfuse} proposed to use CNNs and transformers in separate streams and fuse their outputs. Valanarasu \textit{et al.}~\cite{valanarasu2021medical} proposed a transformer-based axial attention mechanism for 2D medical image segmentation. There are key differences between our model and these efforts: (1) UNETR is tailored for 3D segmentation and directly utilizes volumetric data; (2) UNETR employs the transformer as the main encoder of a segmentation network and directly connects it to the decoder via skip connections, as opposed to using it as an attention layer within the segmentation network (3) UNETR does not rely on a backbone CNN for generating the input sequences and directly utilizes the tokenized patches.

For 3D medical image segmentation, Xie~\textit{et al.}~\cite{xie2021cotr} proposed a framework that utilizes a backbone CNN for feature extraction, a transformer to process the encoded representation and a CNN decoder for predicting the segmentation outputs. Similarly, Wang~\textit{et al.}~\cite{wang2021transbts} proposed to use a transformer in the bottleneck of a 3D encoder-decoder CNN for the task of semantic brain tumor segmentation. In contrast to these approaches, our method directly connects the encoded representation from the transformer to decoder by using skip connections.   

\begin{figure*}[t]
\centering
\includegraphics[width=\textwidth]{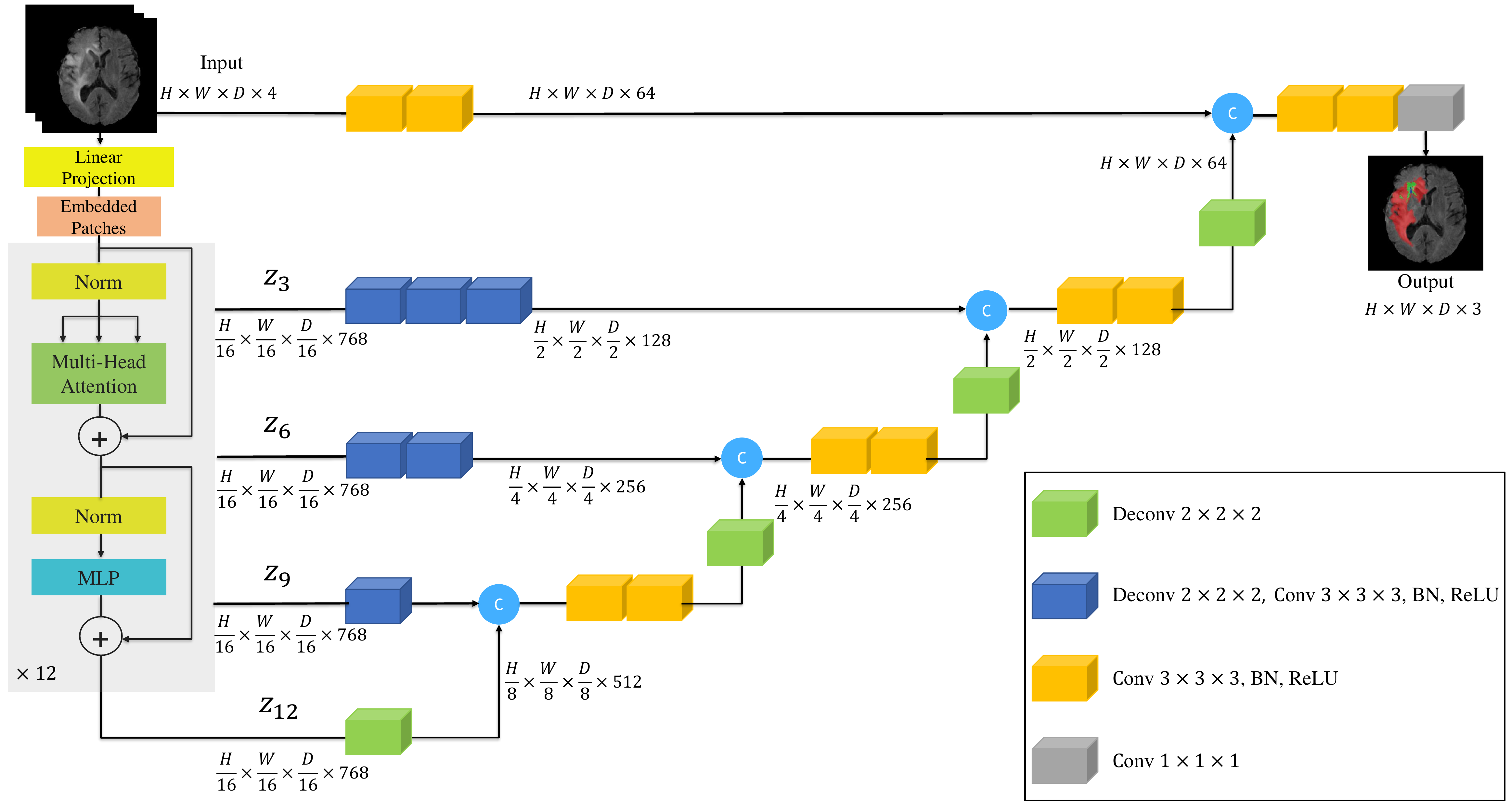}
  \caption{Overview of UNETR architecture. A 3D input volume (e.g. $C=4$ channels for MRI images), is divided into a sequence of uniform non-overlapping patches and projected into an embedding space using a linear layer. The sequence is added with a position embedding and used as an input to a transformer model. The encoded representations of different layers in the transformer are extracted and merged with a decoder via skip connections to predict the final segmentation. Output sizes are given for patch resolution $P=16$ and embedding size $K=768$.}
  \label{fig:pipeline}
\end{figure*}
\section{Methodology}
\label{sec:method}
\subsection{Architecture}
We have presented an overview of the proposed model in Fig.~\ref{fig:pipeline}. UNETR utilizes a contracting-expanding pattern consisting of a stack of transformers as the encoder which is connected to a decoder via skip connections. As commonly used in NLP, the transformers operate on 1D sequence of input embeddings. Similarly, we create a 1D sequence of a 3D input volume $\ \mathbf{x}\in \mathbb{R}^{H\times W\times D\times C}$ with resolution $(H,W,D)$ and $C$ input channels by dividing it into flattened uniform non-overlapping patches $\ \mathbf{x}_{v}\in \mathbb{R}^{N\times (P^{3}.C)}$ where $(P,P,P)$ denotes the resolution of each patch and $N=(H\times W\times D)/P^{3}$ is the length of the sequence. 

Subsequently, we use a linear layer to project the patches into a $K$ dimensional embedding space, which remains constant throughout the transformer layers. In order to preserve the spatial information of the extracted patches, we add a 1D learnable positional embedding $\mathbf{E}_{pos} \in \mathbb{R}^{N\times K}$ to the projected patch embedding $\mathbf{E} \in \mathbb{R}^{(P^{3}.C) \times K}$ according to

\begin{equation}
    \label{eq:trans1}
    \mathbf{z}_{0} = [\mathbf{x}^{1}_{v}\mathbf{E};\mathbf{x}^{2}_{v}\mathbf{E};...;\mathbf{x}^{N}_{v}\mathbf{E}] + \mathbf{E}_{pos},
\end{equation}

Note that the learnable \texttt{[class]} token is not added to the sequence of embeddings since our transformer backbone is designed for semantic segmentation. After the embedding layer, we utilize a stack of transformer blocks~\cite{vaswani2017attention,dosovitskiy2020image} comprising of multi-head self-attention (MSA) and multilayer perceptron (MLP) sublayers according to

\begin{equation}
    \label{eq:trans2}
    \mathbf{z^{\prime}}_{i} = \textnormal{MSA}(\textnormal{Norm}(\mathbf{z}_{i-1})) + \mathbf{z}_{i-1}, \quad i=1 \ldots L,
\end{equation}
\begin{equation}
    \label{eq:trans3}
    \mathbf{z}_{i} = \textnormal{MLP}(\textnormal{Norm}(\mathbf{z^{\prime}}_{i})) + \mathbf{z^{\prime}}_{i}, \quad i=1 \ldots L,
\end{equation}
where Norm() denotes layer normalization~\cite{ba2016layer}, MLP comprises of two linear layers with GELU activation functions, $i$ is the intermediate block identifier, and $L$ is the number of transformer layers. 

A MSA sublayer comprises of $n$ parallel self-attention ($\textnormal{SA}$) heads. Specifically, the $\textnormal{SA}$ block, is a parameterized function that learns the mapping between a query ($\mathbf{q}$) and the corresponding key ($\mathbf{k}$) and value ($\mathbf{v}$) representations in a sequence $\ \mathbf{z}\in \mathbb{R}^{N\times K}$. The attention weights ($\textnormal{A}$) are computed by measuring the similarity between two elements in $\ \mathbf{z}$ and their key-value pairs according to
\begin{equation}
    \label{eq:trans4}
    \textnormal{A} = \textnormal{Softmax}(\frac{{\mathbf{q}\mathbf{k}^\top}}{\sqrt{K_{h}}}),
\end{equation}
where $K_{h}=K/n$ is a scaling factor for maintaining the number of parameters to a constant value with different values of the key $\mathbf{k}$. Using the computed attention weights, the output of $\textnormal{SA}$ for values $\mathbf{v}$ in the sequence $\mathbf{z}$ is computed as
\begin{equation}
    \label{eq:trans6}
    \textnormal{SA}(\mathbf{z}) = \mathbf{A}\mathbf{v},
\end{equation}
Here, $\mathbf{v}$ denotes the values in the input sequence and $K_{h}=K/n$ is a scaling factor. Furthermore, the output of MSA is defined as
\begin{equation}
    \label{eq:trans5}
    \textnormal{MSA}(\mathbf{z}) = [\textnormal{SA}_{1}(\mathbf{z});\textnormal{SA}_{2}(\mathbf{z});...;\textnormal{SA}_{n}(\mathbf{z})] \mathbf{W}_{msa},
\end{equation}
where $\mathbf{W}_{msa}\in \mathbb{R}^{n.K_{h}\times K}$ represents the multi-headed trainable parameter weights. 

Inspired by architectures that are similar to U-Net ~\cite{Ronneberger15}, where features from multiple resolutions of the encoder are merged with the decoder, we extract a sequence representation $\mathbf{z}_{i}$ ($i \in \{3,6,9,12\}$), with size $\frac{H\times W\times D}{P^{3}}\times K$, from the transformer and reshape them into a $\frac{H}{P}\times \frac{W}{P}\times \frac{D}{P}\times K$ tensor. A representation in our definition is in the embedding space after it has been reshaped as an output of the transformer with feature size of $K$ (i.e. transformer's embedding size). Furthermore, as shown in Fig.~\ref{fig:pipeline}, at each resolution we project the reshaped tensors from the embedding space into the input space by utilizing consecutive $3\times3\times3$ convolutional layers that are followed by normalization layers.

At the bottleneck of our encoder (i.e. output of transformer's last layer), we apply a deconvolutional layer to the transformed feature map to increase its resolution by a factor of 2. We then concatenate the resized feature map with the feature map of the previous transformer output (e.g. $\mathbf{z}_{9}$), and feed them into consecutive $3\times3\times3$ convolutional layers and upsample the output using a deconvolutional layer. This process is repeated for all the other subsequent layers up to the original input resolution where the final output is fed into a $1\times 1\times 1$ convolutional layer with a softmax activation function to generate voxel-wise semantic predictions. 

\subsection{Loss Function}
Our loss function is a combination of soft dice loss~\cite{milletari2016v} and cross-entropy loss, and it can be computed in a voxel-wise manner according to
\begin{equation}
\begin{split}
\mathcal{L}(G,Y) &= 1-\frac{2}{J}\sum_{j=1}^{J}\frac{\sum_{i=1}^{I} G_{i,j}Y_{i,j} }{\sum_{i=1}^{I}G^{2}_{i,j}+ \sum_{i=1}^{I}Y^{2}_{i,j}}-\\
& -\frac{1}{I}\sum_{i=1}^{I}\sum_{j=1}^{J} G_{i,j}\log Y_{i,j}.
\end{split}
\label{eq:ce}
\end{equation}
where $I$ is the number of voxels; $J$ is the number of classes; $Y_{i,j}$ and $G_{i,j}$ denote the probability output and one-hot encoded ground truth for class $j$ at voxel $i$, respectively.

\begin{table*}[t]
\scriptsize
\resizebox{\textwidth}{!}{%
\begin{tabular}{lccccccccccccc}
\hline
Methods & \multicolumn{1}{l}{Spl}  
& \multicolumn{1}{l}{RKid} & \multicolumn{1}{l}{LKid} 
& \multicolumn{1}{l}{Gall}  & \multicolumn{1}{l}{Eso} 
& \multicolumn{1}{l}{Liv} & \multicolumn{1}{l}{Sto} 
& \multicolumn{1}{l}{Aor} & \multicolumn{1}{l}{IVC} 
& \multicolumn{1}{l}{Veins}   & \multicolumn{1}{l}{Pan} 
&\multicolumn{1}{l}{AG} & \multicolumn{1}{l}{Avg.} \\ \hline
SETR NUP~\cite{zheng2020rethinking}
& 0.931 & 0.890                        
& 0.897 & 0.652                        
& 0.760 & 0.952                      
& 0.809 & 0.867   
& 0.745 & 0.717
& 0.719 & 0.620
& 0.796
\\ 
SETR PUP~\cite{zheng2020rethinking}  
& 0.929 & 0.893                        
& 0.892 & 0.649                     
& 0.764 & 0.954                        
& 0.822 & 0.869   
& 0.742 & 0.715
& 0.714 & 0.618
& 0.797
\\ 
SETR MLA~\cite{zheng2020rethinking}    
& 0.930 & 0.889                   
& 0.894 & 0.650                   
& 0.762 & 0.953                      
& 0.819 & 0.872  
& 0.739 & 0.720
& 0.716 & 0.614
& 0.796
\\ 
nnUNet~\cite{isensee2021nnu}    
& 0.942 & 0.894                        
& 0.910 & 0.704                        
& 0.723 & 0.948                         
& 0.824 & 0.877   
& 0.782 & 0.720
& 0.680 & 0.616
& 0.802
\\ 
ASPP~\cite{chen2018encoder}                   
& 0.935 & 0.892                        
& 0.914 & 0.689                        
& 0.760 & 0.953                         
& 0.812 & 0.918   
& 0.807 & 0.695
& 0.720 & 0.629
& 0.811  
\\ 
TransUNet~\cite{chen2021transunet}    
& 0.952 & \textbf{0.927}                        
& 0.929 & 0.662                        
& 0.757 & 0.969                        
& 0.889 & 0.920   
& 0.833 & 0.791
& 0.775 & 0.637
& 0.838
\\ 
CoTr w/o CNN encoder~\cite{xie2021cotr}   
& 0.941 & 0.894                        
& 0.909 & 0.705                      
& 0.723 & 0.948                       
& 0.815 & 0.876   
& 0.784 & 0.723
& 0.671 & 0.623
& 0.801
\\ 
CoTr*~\cite{xie2021cotr}     
& 0.943 & 0.924                        
& 0.929 & 0.687                        
& 0.762 & 0.962                        
& 0.894 & 0.914   
& 0.838 & \textbf{0.796}
& \textbf{0.783} & 0.647
& 0.841
\\ 
CoTr~\cite{xie2021cotr}     
& 0.958 & 0.921                    
& 0.936 & 0.700                        
& 0.764 & 0.963                        
& 0.854 & \textbf{0.920}  
& 0.838 & 0.787
& 0.775 & 0.694
& 0.844
\\ 
\textbf{UNETR}  
& \textbf{0.968}    & {0.924}        
& \textbf{0.941}    & \textbf{0.750}        
& \textbf{0.766}    & \textbf{0.971}        
& \textbf{0.913}    & 0.890 
& \textbf{0.847}    & 0.788        
& 0.767    & \textbf{0.741}  
& \textbf{0.856}               
\\ \hline
RandomPatch~\cite{tang2021high}     
& 0.963 & 0.912                        
& 0.921 & 0.749                        
& 0.760 & 0.962                        
& 0.870 & 0.889   
& 0.846 & 0.786
& 0.762 & 0.712
& 0.844
\\ 
PaNN~\cite{zhou2019prior}     
& 0.966 & 0.927                       
& 0.952 & 0.732                        
& 0.791 & 0.973                        
& 0.891 & 0.914   
& 0.850 & 0.805
& 0.802 & 0.652
& 0.854
\\
nnUNet-v2~\cite{isensee2021nnu}     
& 0.972 & 0.924                        
& \textbf{0.958} & 0.780                        
& 0.841 & 0.976                        
& 0.922 & 0.921   
& 0.872 & 0.831
& 0.842 & 0.775
& 0.884
\\ 
nnUNet-dys3~\cite{isensee2021nnu}     
& 0.967 & 0.924                        
& 0.957 & 0.814                        
& 0.832 & 0.975                        
& 0.925 & 0.928   
& 0.870 & 0.832
& \textbf{0.849} & 0.784
& 0.888
\\ 
\textbf{UNETR}     
& \textbf{0.972} & \textbf{0.942}                        
& 0.954 & \textbf{0.825}                        
& \textbf{0.864} & \textbf{0.983}                        
& \textbf{0.945} & \textbf{0.948}   
& \textbf{0.890} & \textbf{0.858}
& 0.799 & \textbf{0.812}
& \textbf{0.891}
\\ \hline
\end{tabular}%
}
\\
\caption{Quantitative comparisons of segmentation performance in BTCV test set. Top and bottom sections represent the benchmarks of Standard and Free Competitions respectively. Our method is compared against current state-of-the-art models. All SETR~\cite{zheng2020rethinking} baselines use ViT-B-16~\cite{dosovitskiy2020image} backbone. Note: Spl: spleen, RKid: right kidney, LKid: left kidney, Gall: gallbladder, Eso: esophagus, Liv: liver, Sto: stomach, Aor: aorta IVC: inferior vena cava, Veins: portal and splenic veins, Pan: pancreas, AG: adrenal gland. All results obtained from BTCV leaderboard.}
\label{tab:BTCV}
\end{table*}

\section{Experiments}

\subsection{Datasets}
\label{sec:datasets}

To validate the effectiveness of our method, we utilize BTCV~\cite{landman2015miccai} and MSD~\cite{simpson2019large} datasets for three different segmentation tasks in CT and MRI imaging modalities.

\textbf{BTCV (CT): } The BTCV dataset~\cite{landman2015miccai} consists of 30 subjects with abdominal CT scans where 13 organs were annotated by interpreters under supervision of clinical radiologists at Vanderbilt University Medical Center. Each CT scan was acquired with contrast enhancement in portal venous phase and consists of $80$ to $225$ slices with $512\times 512$ pixels and slice thickness ranging from $1$ to $6$ $mm$. Each volume has been pre-processed independently by normalizing the intensities in the range of [-1000,1000] HU to [0,1]. All images are resampled into the isotropic voxel spacing of 1.0 $mm$ during pre-processing. The multi-organ segmentation problem is formulated as a 13 class segmentation task with 1-channel input.

\begin{figure*}[t]
\centering
\includegraphics[width=0.95\textwidth]{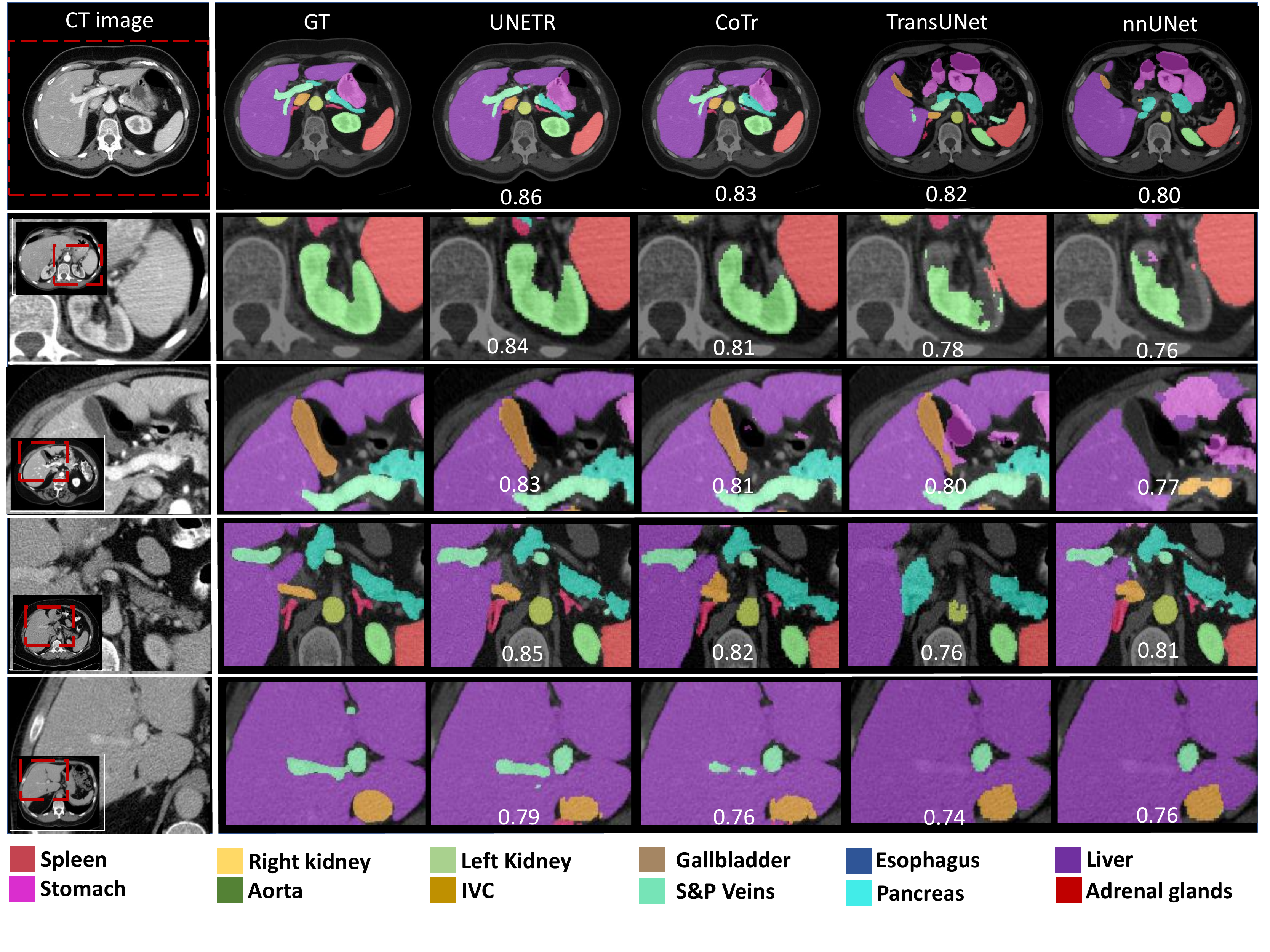}
\caption{Qualitative comparison of different baselines in BTCV cross-validation. The first row shows a complete representative CT slice. We exhibit four zoomed-in subjects (row 2 to 5), where our method shows visual improvement on segmentation of kidney and spleen (row 2), pancreas and adrenal gland (row 3), gallbladder (row 4) and portal vein (row 5). The subject-wise average Dice score is shown on each sample.}
\label{fig:qualitative}
\end{figure*}

\begin{table*}[t]
\centering
\scriptsize
\resizebox{\textwidth}{!}{%
\begin{tabular}{lccllllllll}
\hline
\multicolumn{1}{l}{Task/Modality}  
& \multicolumn{2}{c}{Spleen Segmentation (CT)} & \multicolumn{8}{c}{Brain tumor Segmentation (MRI)} \\
\multicolumn{1}{l}{Anatomy}  
& \multicolumn{2}{c}{Spleen} & \multicolumn{2}{c}{WT}
& \multicolumn{2}{c}{ET} & \multicolumn{2}{c}{TC}
& \multicolumn{2}{c}{All} \\ 
\cmidrule(r){2-3} 
\cmidrule(r){4-5} 
\cmidrule(r){6-7} 
\cmidrule(r){8-9}
\cmidrule(r){10-11}
Metrics & \multicolumn{1}{c}{Dice}  & \multicolumn{1}{c}{HD95}
& Dice  & HD95
& Dice  & HD95
& Dice  & HD95
& Dice  & HD95 

\\ \hline
UNet~\cite{Ronneberger15}    
& 0.953 & 4.087               
& 0.766 & 9.205  
& 0.561 & 11.122   
& 0.665 & 10.243 
& 0.664 & 10.190  
\\
AttUNet~\cite{oktay2018attention}    
& 0.951 & 4.091               
& 0.767 & 9.004  
& 0.543 & 10.447   
& 0.683 & 10.463 
& 0.665 & 9.971  
\\ 
SETR NUP~\cite{zheng2020rethinking}    
& 0.947 & 4.124               
& 0.697 & 14.419  
& 0.544 & 11.723   
& 0.669 & 15.192 
& 0.637 & 13.778  
\\
SETR PUP~\cite{zheng2020rethinking}    
& 0.949 & 4.107               
& 0.696 & 15.245  
& 0.549 & 11.759   
& 0.670 & 15.023 
& 0.638 & 14.009  
\\
SETR MLA~\cite{zheng2020rethinking}    
& 0.950 & 4.091               
& 0.698 & 15.503  
& 0.554 & 10.237   
& 0.665 & 14.716 
& 0.639 & 13.485  
\\
TransUNet~\cite{chen2021transunet}    
& 0.950 & 4.031               
& 0.706 & 14.027  
& 0.542 & 10.421   
& 0.684 & 14.501 
& 0.644 & 12.983  
\\
TransBTS~\cite{wang2021transbts}     
& - & -               
& 0.779 & 10.030  
& 0.574 & 9.969   
& 0.735 & 8.950 
& 0.696 & 9.650  
\\ 
CoTr w/o CNN encoder~\cite{xie2021cotr}   
& 0.946 & 4.748               
& 0.712 & 11.492  
& 0.523 & 9.592   
& 0.698 & 12.581 
& 0.6444 & 11.221  
\\ 
CoTr~\cite{xie2021cotr}     
& 0.954 & 3.860               
& 0.746 & 9.198  
& 0.557 & 9.447   
& 0.748 & 10.445 
& 0.683 & 9.697  
\\ 
\textbf{UNETR}  
& \textbf{0.964}    & \textbf{1.333}        
& \textbf{0.789}    & \textbf{8.266}  
& \textbf{0.585}    & \textbf{9.354} 
& \textbf{0.761}    & \textbf{8.845} 
& \textbf{0.711}    & \textbf{8.822} 
\\ \hline
\end{tabular}%
}
\\
\caption{Quantitative comparisons of the segmentation performance in brain tumor and spleen segmentation tasks of the MSD dataset. WT, ET and TC denote Whole Tumor, Enhancing tumor and Tumor Core sub-regions respectively.}
\label{tab:msd}
\end{table*}


\textbf{MSD (MRI/CT): } For the brain tumor segmentation task, the entire training set of 484 multi-modal multi-site MRI data (FLAIR, T1w, T1gd, T2w) with ground truth labels of gliomas segmentation necrotic/active tumor and oedema is utilized for model training. The voxel spacing of MRI images in this tasks is $1.0 \times 1.0 \times 1.0$ $mm^3$. The voxel intensities are pre-processed with z-score normalization. The problem of brain tumor segmentation is formulated as a 3 class segmentation task with 4-channel input.

For the spleen segmentation task, 41 CT volumes with spleen body annotation are used.
The resolution/spacing of volumes in task 9 ranges from $0.613 \times 0.613 \times 1.50$ $mm^3$ to $0.977 \times 0.977 \times 8.0$ $mm^3$. All volumes are re-sampled into the isotropic voxel spacing of 1.0 $mm$ during pre-processing. The voxel intensities of the images are normalized to the range $\left[0,1\right]$ according to 5th and 95th percentile of overall foreground intensities. Spleen segmentation is formulated as a binary segmentation task with 1-channel input. For multi-organ and spleen segmentation tasks, we randomly sample the input images with volume sizes of $[96,96,96]$. For brain segmentation task, we randomly sample the input images with volume sizes of $[128,128,128]$. For all experiments, the random patches of foreground/background are sampled at ratio $1:1$. 

\subsection{Evaluation Metrics}
We use Dice score and 95\% Hausdorff Distance (HD) to evaluate the accuracy of segmentation in our experiments. For a given semantic class, let $G_{i}$ and $P_{i}$ denote the ground truth and prediction values for voxel $i$ and $G'$ and $P'$ denote ground truth and prediction surface point sets respectively. The Dice score and HD metrics are defined as
\begin{equation}
\textrm{Dice}(G,P)= \frac{2\sum_{i=1}^{I} G_{i}P_{i} }{\sum_{i=1}^{I}G_{i}+ \sum_{i=1}^{I}P_{i}},
\label{eq:dice_score}
\end{equation}
\begin{equation}
\begin{split}
\textrm{HD}(G',P')=\max \{{\max _{g' \in G'} \min _{p' \in P'} } \|g'-p'\|, \\
\max _{p' \in P'} \min_{g' \in G'} \|p'-g'\| \}.
\end{split}
\label{eq:hd_score}
\end{equation}
The 95\% HD uses the 95th percentile of the distances between ground truth and prediction surface point sets. As a result, the impact of a very small subset of outliers is minimized when calculating HD.

\begin{figure*}[t]
\centering
\includegraphics[width=0.95\textwidth]{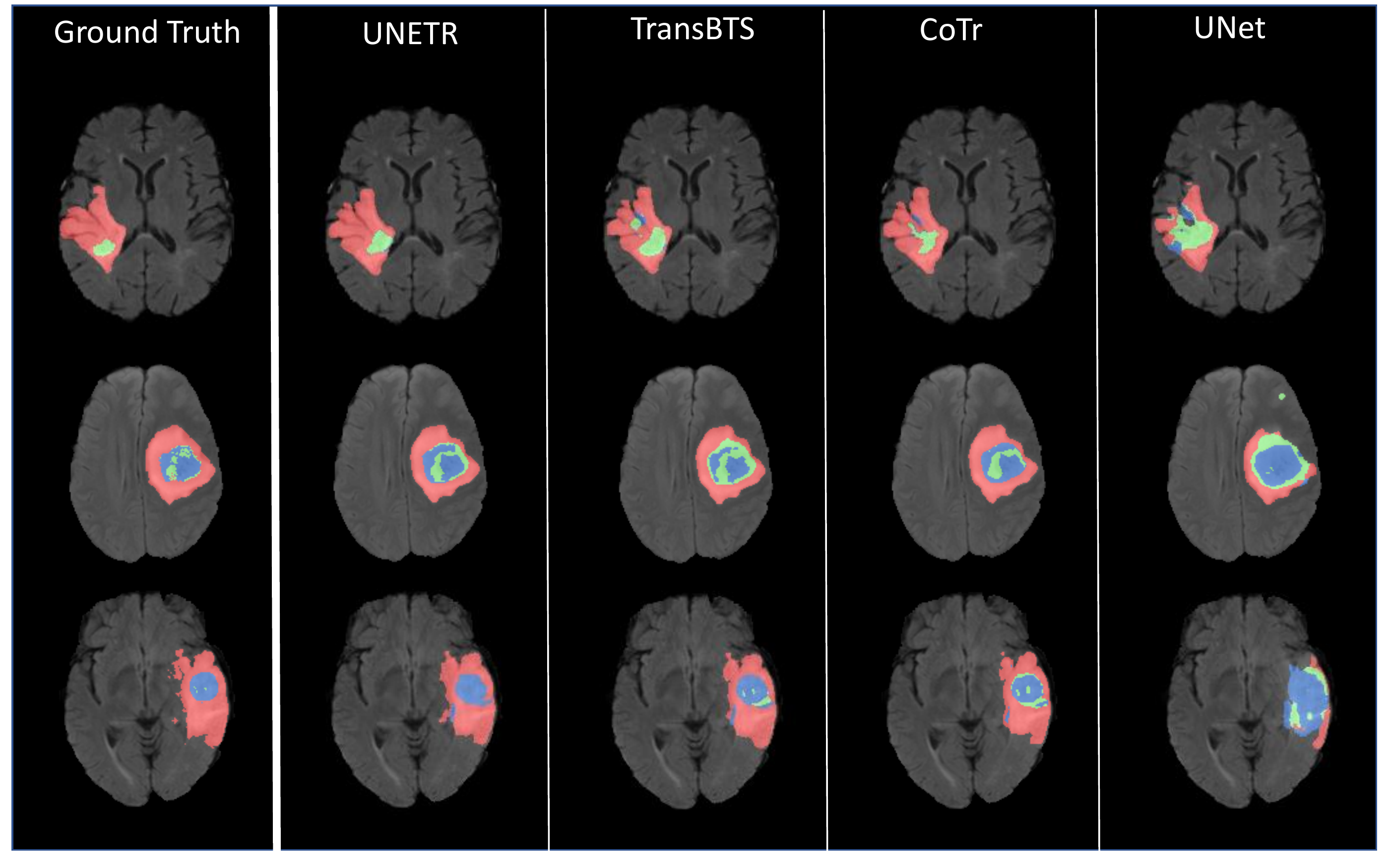}
  \caption{UNETR effectively captures the fine-grained details in segmentation outputs. The Whole Tumor (WT) encompasses a union of red, blue and green regions.  The Tumor Core (TC) includes the union of red and blue regions. The Enhancing Tumor core (ET) denotes the green regions.}
  \label{fig:brats}
\end{figure*}

\subsection{Implementation Details}

We implement UNETR in PyTorch\footnote{\href{http://pytorch.org/}{http://pytorch.org/}} and MONAI\footnote{\href{https://monai.io/}{https://monai.io/}}. The model was trained using a NVIDIA DGX-1 server.
All models were trained with the batch size of $6$, using the AdamW optimizer~\cite{loshchilov2017decoupled} with initial learning rate of $0.0001$ for $20,000$ iterations. For the specified batch size, the average training time was 10 hours for 20,000 iterations. Our transformer-based encoder follows the ViT-B16~\cite{dosovitskiy2020image} architecture with $L=12$ layers, an embedding size of $K=768$. We used a patch resolution of $16\times 16\times 16$. For inference, we used a sliding window approach with an overlap portion of 0.5 between the neighboring patches and the same resolution as specified in Sec.~\ref{sec:datasets}. We did not use any pre-trained weights for our transformer backbone (e.g. ViT on ImageNet) since it did not demonstrate any performance improvements. For BTCV dataset, we have evaluated our model and other baselines in the Standard and Free Competitions of its leaderboard. Additional data from the same cohort was used for the Free Competition increasing the number of training cases to 80 volumes. For all experiments, we employed five-fold cross validation with a ratio of 95:5. In addition, we used data augmentation strategies such as random rotation of 90, 180 and 270 degrees, random flip in axial, sagittal and coronal views and random scale and shift intensity. We used ensembling to fuse the outputs of models from four different five-fold cross-validations. For brain and spleen segmentation tasks in MSD dataset, we split the data into training, validation and test with a ratio of 80:15:5.

\subsection{Quantitative Evaluations}
UNETR outperforms the state-of-the-art methods for both Standard and Free Competitions on the BTCV leaderboard. As shown in Table~\ref{tab:BTCV}, in the Free Competition, UNETR achieves an overall average Dice score of $0.899$ and outperforms the second, third and fourth top-ranked methodologies by $1.238\%$, $1.696\%$ and $5.269\%$ respectively. 

In the Standard Competition, we compared the performance of UNETR against CNN and transformer-based baselines. UNETR achieves a new state-of-the-art performance with an average Dice score of $85.3\%$ on all organs. Specifically, on large organs, such as spleen, liver and stomach, our method outperforms the second best baselines by $1.043\%$, $0.830\%$ and $2.125\%$ respectively,in terms of Dice score. Furthermore, in segmentation of small organs, our method significantly outperforms the second best baselines by $6.382\%$ and $6.772\%$ on gallbladder and adrenal glands respectively, in terms of Dice score.
 
In Table~\ref{tab:msd}, we compare the performance of UNETR against CNN and transformer-based methodologies for brain tumor and spleen segmentation tasks on MSD dataset. For brain segmentation, UNETR outperforms the closest baseline by $1.5\%$ on average over all semantic classes. In particular, UNETR performs considerably better in segmenting tumor core (TC) subregion. Similarly for spleen segmentation, UNETR outperforms the best competing methodology by least $1.0\%$ in terms of Dice score.

\subsection{Qualitative Results}

Qualitative multi-organ segmentation comparisons are presented in Fig.~\ref{fig:qualitative}. UNETR shows improved segmentation performance  for abdomen organs. Our model's capability of learning long-range dependencies is evident in row 3 (from the top), in which nnUNet confuses liver with stomach tissues, while UNETR successfully delineates the boundaries of these organs. In Fig.~\ref{fig:qualitative}, rows 2 and 4 demonstrate a clear detection of kidney and adrenal glands against surrounding tissues, which indicate that UNETR captures better spatial context. In comparison to 2D transformer-based models, UNETR exhibits higher boundary segmentation accuracy as it accurately identifies the boundaries between kidney and spleen. This is evident for gallbladder in row 2, liver and stomach in row 3, and portal vein against liver in row 5. In Fig.~\ref{fig:brats}, we present qualitative segmentation comparisons for brain tumor segmentation on the MSD dataset. Specifically, our model demonstrates better performance in capturing the fine-grained details of tumors. 

\section{Discussion}
Our experiments in all datasets demonstrate superior performance of UNETR over both CNN and transformer-based segmentation models. Specifically, UNETR achieves better segmentation accuracy by capturing both global and local dependencies. In qualitative comparisons, this is illustrated in various cases in which UNETR effectively captures long-range dependencies (e.g. accurate segmentation of the pancreas tail in Fig.~\ref{fig:qualitative}).

Moreover, the segmentation performance of UNETR on the BTCV leaderboard demonstrates new state-of-the-art benchmarks and validates its effectiveness. Specifically for small anatomies, UNETR outperforms both CNN and transformer-based models. Although 3D models already demonstrate high segmentation accuracy for small organs such as gallbladder, adrenal glands, UNETR can still outperform the best competing model by a significant margin (See Table~\ref{tab:BTCV}). This is also observed in Fig.~\ref{fig:qualitative}, in which UNETR has a significantly better segmentation accuracy for left and right adrenal glands, and UNETR is the only model to correctly detect branches of the adrenal glands. For more challenging tissues, such as gallbladder in row 4 and portal vein in row 5, which have low contrast with the surrounding liver tissue, UNETR is still capable of segmenting clear connected boundaries.

\section{Ablation}
\paragraph{Decoder Choice}
In Table~\ref{tab:decoder}, we evaluate the effectiveness of our decoder by comparing the performance of UNETR with other decoder architectures on two representative segmentation tasks from MRI and CT modalities. In these experiments, we employ the encoder of UNETR but replaced the decoder with 3D counterparts of Naive UPsampling (NUP), Progressive UPsampling (PUP) and MuLti-scale Aggregation (MLA)~\cite{zheng2020rethinking}. We observe that these decoder architectures yield sub-optimal performance, despite MLA marginally outperforming both NUP and PUP. For brain tumor segmentation, UNETR outperforms its variants with MLA, PUP and NUP decoders by $2.7\%$, $4.3\%$ and $7.5\%$ on average Dice score. Similarly, for spleen segmentation, UNETR outerforms MLA, PUP and NUP by $1.4\%$, $2.3\%$ and $3.2\%$. 

\begin{table}[]
 \centering
\begin{tabular}{lrrrrr}
\hline
Organ   & \multicolumn{1}{l}{Spleen}         & \multicolumn{4}{c}{Brain} \\
\cmidrule(r){3-6}
Decoder & Spleen                             & WT                                 & ET                                 & TC                                 & All                                \\ \hline
NUP     & 0.932                              & 0.721                              & 0.527                              & 0.660                              & 0.636                              \\
PUP     & 0.941                              & 0.749                              & 0.558                              & 0.698                              & 0.668                              \\
MLA     & 0.950                              & 0.757                              & 0.563                              & 0.732                              & 0.684                             \\
\textbf{UNETR}   & \textbf{0.964} &\textbf{0.789} & \textbf{0.585} &\textbf{0.761} & \textbf{0.711}
\\
\hline
\\
\end{tabular}%

\caption{Effect of the decoder architecture on segmentation performance. NUP, PUP and MLA denote Naive UpSampling, Progressive UpSampling and Multi-scale Aggregation.}
\label{tab:decoder}
\end{table}


\paragraph{Patch Resolution}
A lower input patch resolution leads to a higher sequence length, and therefore higher memory consumption, since it is inversely correlated to the cube of the resolution. As shown in Table~\ref{tab:patch}, our experiments demonstrate that decreasing the resolution leads to consistently improved performance. Specifically, decreasing the patch resolution from 32 to 16 improves the performance by $1.1\%$ and $0.8\%$ in terms of average Dice score in spleen and brain segmentation tasks respectively. We did not experiment with lower resolutions due to memory constraints.

\begin{table}[]
 \centering
\begin{tabular}{lrrrrr}
\hline
Organ   & \multicolumn{1}{l}{Spleen}         & \multicolumn{4}{c}{Brain} \\
\cmidrule(r){3-6}
Resolution & Spleen                             & WT                                 & ET                                 & TC                                 & All                                \\ \hline
32     & 0.953                              & 0.776                              & 0.579                              & 0.756                              & 0.703                              \\
16   & \textbf{0.964} &\textbf{0.789} & \textbf{0.585} &\textbf{0.761} & \textbf{0.711}
\\
\hline
\\
\end{tabular}%
\caption{Effect of patch resolution on segmentation performance.}
\label{tab:patch}
\end{table}

\begin{table}[]
 \centering
\resizebox{\columnwidth}{!}{
\begin{tabular}{lrrr}
\hline
Models   & \#Params (M)  & FLOPs (G) & Inference Time (s) \\
\hline
nnUNet \cite{isensee2021nnu}   & 19.07   & 412.65  & 10.28   \\ 
CoTr \cite{xie2021cotr}   & 46.51   & 399.21  & 19.21   \\ 
TransUNet \cite{chen2021transunet}     & 96.07 & 48.34 & 26.97  \\
ASPP \cite{deeplabv3plus2018}   & 47.92   & 44.87  & 25.47   \\ 
SETR \cite{zheng2020rethinking}     & 86.03 & 43.49 & 24.86 \\
\bf{UNETR}   & 92.58 & 41.19 & 12.08\\
\hline
\\
\end{tabular}%
}
\caption{ Comparison of number of parameters, FLOPs and averaged inference time for various models in BTCV experiments.}
\label{tab:complexity}
\end{table}


\paragraph{Model and Computational Complexity}

In Table~\ref{tab:complexity}, we present number of FLOPs, parameters and averaged inference time of the models in BTCV benchmarks. Number of FLOPs and inference time are calculated based on an input size of $96\times96\times96$ and using a sliding window approach. According to our benchmarks, UNETR is a moderate-sized model with 92.58M parameters and 41.19G FLOPs. For comparison, other transformer-based methods such as CoTr~\cite{xie2021cotr}, TransUNet~\cite{chen2021transunet} and SETR~\cite{zheng2020rethinking} have 46.51M, 96.07M and 86.03M parameters and 399.21G, 48.24G and 43.49G FLOPs, respectively. UNETR shows comparable model complexity while outperforming these models by a large margin in BTCV benchmarks. CNN-based segmentation models of nnUNet~\cite{isensee2021nnu} and ASPP~\cite{chen2018encoder} have 19.07M and 47.92M parameters and 412.65G and 44.87G FLOPs, respectively. Similarly, UNETR outperforms these CNN-based models while having a moderate model complexity. In addition, UNETR has the second lowest averaged inference time after nnUNet~\cite{isensee2021nnu} and is significantly faster than transformer-based models such as SETR~\cite{zheng2020rethinking}, TransUNet~\cite{chen2021transunet} and CoTr~\cite{xie2021cotr}.




\section{Conclusion}

This paper introduces a novel transformer-based architecture, dubbed as UNETR, for semantic segmentation of volumetric medical images by reformulating this task as a 1D sequence-to-sequence prediction problem. We proposed to use a transformer encoder to increase the model's capability for learning long-range dependencies and effectively capturing global contextual representation at multiple scales. 

We validated the effectiveness of UNETR on different volumetric segmentation tasks in CT and MRI modalities. UNETR achieves new state-of-the-art performance in both Standard and Free Competitions on the BTCV leaderboard for the multi-organ segmentation and outperforms competing approaches for brain tumor and spleen segmentation on the MSD dataset. In conclusion, UNETR has shown the potential to effectively learn the critical anatomical relationships represented in medical images. The proposed method could be the foundation for a new class of transformer-based segmentation models in medical images analysis.
{\small
\bibliographystyle{ieee_fullname}
\bibliography{egbib}
}

\end{document}